\documentclass[usenatbib,usegraphicx]{mn2e}
\usepackage{subfigure}
\usepackage{longtable}
\usepackage{graphicx}
\usepackage{graphics}
\usepackage{keyval}
\usepackage{trig}
\def\beq{\begin{equation}}
\def\eeq{\end{equation}}
\def\bey{\begin{eqnarray}}
\def\eey{\end{eqnarray}}

\def\msun{M_\odot}

\def\lsim{\mathrel{\raise.3ex\hbox{$<$\kern-.75em\lower1ex\hbox{$\sim$}}}}
\def\gsim{\mathrel{\raise.3ex\hbox{$  $\kern-.75em\lower1ex\hbox{$\sim$}}}}

\def\lsun{L_\odot}
\def\kms{\, {\rm km \, s}^{-1} }

\title{The velocity  distribution of SDSS satellites in MOND}
\author[Angus et al.]{G. W. Angus$^{1}$\thanks{email:
gwa2@st-andrews.ac.uk}, B. Famaey$^{2}$\thanks{email:
bfamaey@ulb.ac.be}, O. Tiret$^{3}$, F. Combes$^{3}$, H.S. Zhao$^{1}$\\
$^{1}$SUPA, School of Physics and Astronomy, University of St. Andrews, Scotland KY16 9SS\\
$^{2}$Institut d'Astronomie et d'Astrophysique, Universit\'e Libre  
de Bruxelles, CP 226, Bvd du Triomphe, B-1050, Bruxelles, Belgium\\
$^{3}$Observatoire de Paris, LERMA, 61 Av. de l'Observatoire, 75014 Paris, France}
\begin{document}

\date{Accepted ... Received ... ; in original form ...}

\pagerange{\pageref{firstpage}--\pageref{lastpage}} \pubyear{2007}

\maketitle

\label{firstpage}
\begin{abstract}
The recent SDSS measured velocity distribution of satellite galaxies has been modelled in the context of MOND.
We show that even when the extra constraint of adhering to the projected satellite number density profile is added, the two line of sight (los) velocity dispersion profiles presented in Klypin \& Prada (2007) can be matched simply with a radially varying anisotropy. Interestingly, the anisotropies required to fit the los velocity dispersions are remarkably similar to the anisotropies generated by dissipationless collapse simulations in MOND. The mass-to-light ratios of the two host galaxies used are sensible and positivity of the distribution function is satisfied.
\end{abstract}


\section{Introduction}
\protect\label{sec:intr}

Ever since Milgrom proposed the Modified Newtonian Dynamics (MOND) in a series of groundbreaking papers (Milgrom 1983abc; Bekenstein \& Milgrom 1984) there have been attempts to falsify the theory using myriad observations and techniques. Many have suffered from poor data, which after being re-analysed was in total agreement with MOND (Milgrom 1995) or they sank because the MOND analysis was done poorly (see Milgrom \& Sanders 2003). However some issues stand because they are legitimate concerns; most notably the missing mass in clusters (Aguirre et al. 2001; Sanders 2003, 2007; Clowe et al. 2006; Angus et al. 2007a, Angus, Famaey \& Buote 2007b). Thanks to the rapid development of MOND N-body codes (Ciotti et al. 2006; Tiret \& Combes 2007), issues of galaxy stability are being treated and merging timescales of galaxy pairs have been shown to be borderline high in recent simulations by Nipoti et al. (2007b).

Generally, it is forlorn to attack MOND at the galaxy scale because it so outperforms CDM even with zero free parameters, the tidal dwarf galaxies observed by Bournaud et al. (2007) being a great example of this (Gentile et al. 2007, Milgrom 2007), as well as the tight correlation between dark and luminous mass which is inferred under the dark matter paradigm (McGaugh 2005). For a review of recent successes of MOND see Bekenstein (2006). However, the Sloan Digital Sky Survey (SDSS) brings a new dimension to our ability to test MOND. With such a vast archive of galaxies, and thanks to its piercing magnitude range, Klypin \& Prada (2007, hereafter KP07) were able to generate the satellite line of sight dispersions for a narrow range of galaxy luminosity. In this small luminosity range, the line of sight (los) dispersions of many satellites were stacked together to essentially create a mock galaxy group with a los dispersion known over a range of radii (50-400kpc). Since there is no evidence for dark matter in small MONDian groups (Milgrom 1998, 2002; but see Buote \& Canizares 1994, 1996, Buote et al. 2002 and Angus et al. 2007b) one should expect the velocity dispersions calculated from the MOND gravity of the host galaxy in the Jeans equations to coincide with the observed ones. In their recent preprint KP07 claimed that MOND ``dramatically fails" to reproduce the falling velocity dispersions. However, their Jeans modelling is based on fairly crude assumptions. Here we re-examine this issue with detailed models.


In \S2 we explain our method for correctly solving the Jeans equation in MOND and then in \S3 we discuss how we proceeded to fit the data for the two representative host galaxies discussed in KP07. 

\section{The Jeans equations in MOND}
KP07 chose red galaxies whose geometry is mostly spherical as host galaxies for the satellite distribution. Then, irrespective of the gravitational theory, to calculate the radial velocity dispersions $\sigma_r(r)$ of an equilibrium population in a given spherically symmetric gravitational field, we must solve the Jeans equation
\beq
\protect\label{eqn:jeans}
{d \over dr}\sigma_r^2(r) + {\gamma(r) \over r}\sigma_r^2(r) = -g(r), \qquad \gamma(r)=\alpha(r)+2\beta(r)
\eeq
where g(r) is the modulus of the internal gravity. The function $\beta(r)=1- \sigma_t^2(r) / 2\sigma_r^2(r)$ is the anisotropy parameter, where $\sigma_t$ is the 2-component tangential dispersion and $\sigma_r$ is the radial dispersion. The function $\alpha(r)=d\ln p(r) / d \ln r$ is the logarithmic gradient of the 3-dimensional number density profile $p(r)$ of satellites. Actually, in the case of satellites, $p(r)$ is rather their probability distribution in configuration space given the few numbers of observed satellites per host galaxy.
However, in the case of satellites, the mass density $\rho(r)$ that will be used in the Poisson equation is {\it not} the same as the number density (or probability distribution) $p(r)$ of satellites. Satellites will just be considered as a population of test particles in equilibrium in a dominating external potential generated by the host galaxy. In this respect, the velocity dispersions expected for planetary nebulae at large radii (e.g., Douglas et al. 2007) will be much lower than that of the satellites. 
The comparison of the large-scale ($\sim$ 300kpc) to small-scale ($\sim$ 30kpc) 
velocity dispersion and anisotropy will be explored 
in detail for well studied galaxies in a future paper (Tiret et al. in prep.). Numerical simulations of elliptical galaxy formation through mergers in the MOND regime will also be reported in future works. 

We note that the parameter $\alpha(r)$ for the satellite number density is taken to be constant in KP07, but clearly decreases from the observed surface density (Fig.1 of KP07). For this reason, we hereafter consider $\alpha$ and $\beta$ (and therefore $\gamma$) to be functions of $r$. We take a simple double-power law profile such that the probability density goes like (Zhao 1996)
\beq
\protect\label{eqn:gar}
p(r) \propto \left(1+{r \over r_{\alpha}}\right)^{\alpha_o}
\eeq
where $\alpha_o$ is the asymptotic slope of the probability density and $r_{\alpha}$ is the break radius of the satellite distribution. It is simple to show that the logarithmic density slope $\alpha(r)={d \ln p \over d \ln r}$ is given by
\beq
\protect\label{eqn:alpha}
\alpha(r)=\alpha_o{r \over r+r_{\alpha}}.
\eeq
Similarly, we can define the anisotropy $\beta(r)$ as
\beq
\protect\label{eqn:beta}
\beta(r)={r-r_{\beta} \over r+r_{\alpha}}
\eeq
Where $r_{\beta}$ is the radius where orbits are isotropic. We use the standard technique of solving first order, linear ODEs and multiply both sides by an integrating factor. From there it can be solved numerically at any given radius, $r_o$ via the equation
\bey
\protect\label{eqn:hub}
\nonumber \sigma_r^2(r_o) &=& r_o^{2r_{\beta}/r_{\alpha}}(r_o+r_{\alpha})^{-\alpha_o-2-2r_{\beta}/r_{\alpha}} \\ &\times& \int_{r_o}^{\infty} r^{-2r_{\beta}/r_{\alpha}}(r+r_{\alpha})^{\alpha_o+2+2r_{\beta}/r_{\alpha}} g(r)\,dr.
\eey
For the integration limits we set $\sigma_r(100Mpc)=0$ which physically must be true. 

\subsection{The Poisson equation and the external field effect}

Now, we need to compute the gravity $g(r)$ from the density of the host galaxy and the adopted gravitational theory. In MOND, the Poisson equation reads \beq
\label{Poisson} -\nabla.[\mu(x) {\bf g} ] = 4 \pi G \rho,
\qquad x \equiv {g \over a_o}, \eeq where $a_o=3.6(\kms)^2 / pc$ is the MOND acceleration constant, and the $\mu$-function is chosen, from Famaey \& Binney (2005), to be $\mu(x)=x/(1+x)$.
Here, given that satellites are located far away from the bulk of the mass of the host galaxy, we model the density as a point mass.

When analyzing the internal gravity of a system at large radii, as in the case of satellites, a role may be played by the ``external field effect" (EFE) linked with the breaking of the Strong Equivalence Principle inherent to any acceleration-based modification of gravity.


However, for these isolated galaxies (unlike the Milky Way), we have no information on the proximity and masses of nearby massive galaxies. In fact, since these galaxies are stacked together, the individual external field would be different from host to host, therefore, it makes no sense to include a single value for it. If there is indeed an EFE, then to remain consistent with the los dispersions, it must in general be less than $a_o/100$. Here we do not consider the EFE.


The internal gravity $g$ of an isolated spherical galaxy in MOND is then determined by
\beq
\protect\label{eqn:mond}
g \mu \left(g/a_o\right)=GMr^{-2}
\eeq
where $g$ is the internal gravitational field of the system which we are interested in, and $M$ is the mass of the host galaxy. 



\section{Results}
Once Eqs.(\ref{eqn:hub}) and (\ref{eqn:mond}) are solved, the radial dispersions must be cast into line-of-sight dispersion in order to compare with the SDSS data. The projected number density $\Sigma(R)=2 \int_R^{\infty} r p(r) {dr \over \sqrt{r^2-R^2}}$ (see Binney \& Mamon 1982) is fitted with $r_{\alpha}=40kpc$ and $\alpha_o=-3.1$ and is shown in our Fig.1 along with the data points given by KP07.

\begin{table} 
\begin{tabular}{|c|c|c|c|c|c|} 
${\rm Galaxy}$ &Mass [${\rm 10^{11}\msun }]$ & $ M/L_g $&$\alpha_o$&$ r_{\alpha}$ [kpc]&$r_{\beta}$ [kpc] \\
1 & 0.8 & 1.9-3.3 & -3.1 & 40 & 63 \\
2 & 2.8 & 4.2-6.6 & -3.1 & 40 & 63 \\
\end{tabular} 
\medskip 
\caption{Shows the parameters used in the MOND simultaneous fitting of the projected number density of satellite galaxies and los dispersions presented by KP07.} 
\end{table} 

This left us with only the galaxy mass, and $\beta(r)$ as free parameters to vary in an attempt to fit the satellite los velocity dispersion of hosts in both magnitude binnings (as described in their Fig.2) by keeping the mass as close to the two representative galaxies given by KP07. The higher we push the galaxy masses, the easier it becomes to fit the los dispersions, however, our $M$ has to be physically consistent with what is expected for the host galaxies i.e. M/L of a few solar units.

The two representative galaxies arise because host galaxies over two ranges of magnitude have their velocity dispersions binned together and a representative mass for that particular range of magnitudes is chosen. The two ranges are g-band luminosities between 2.4 and 4.2$\times10^{10}\lsun$ for galaxy 1 and between 4.2 and 6.6$\times10^{10}\lsun$ for galaxy 2. KP07 take $M/L_g \sim 2-3$ for galaxy 1 and $M/L_g \sim 3-5$ for galaxy 2. Galaxy 2 actually corresponds to very red galaxies and so there is considerable potential for variability of the chosen mass not just because the binning of galaxies relates to luminosities that vary around their mean by $\sim$25\%, but also because of the uncertainty in $M/L_g$. 

\begin{figure*}
\includegraphics[angle=0,width=17.2cm]{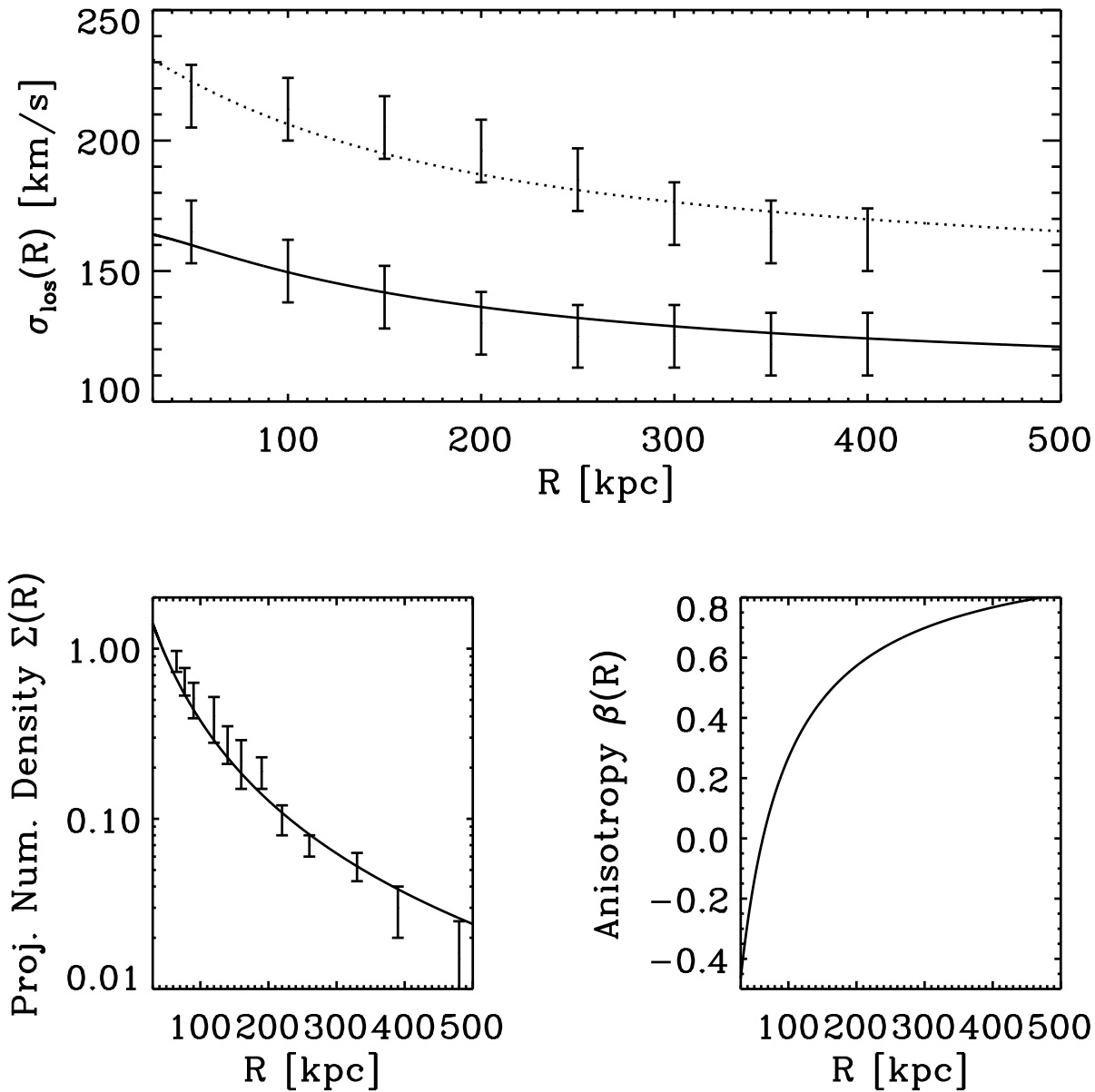}
\caption{{\it Top panel}: The expected los velocity dispersion profiles in MOND for the two host galaxies along with the observed velocity dispersion profiles with 1$\sigma$ errors. The solid linetype is for galaxy 1: $M=8.0\times 10^{10}\msun$ for $r_{\alpha}=40kpc$, $\alpha_o=-3.1$, and $r_{\beta}=63kpc$. The dashed linetype is for galaxy 2: $M=2.8\times 10^{11}\msun$ and all other parameters are the same as for galaxy 1. {\it Bottom left}: The surface density profile we used for the fits to the velocity dispersions. The fit involves choosing $r_{\alpha}=40kpc$ and $\alpha_o=-3.1$ for Eq.\ref{eqn:gar} and then integrating along the line of sight. The data points are the observed satellite surface densities from KP07 {\it Bottom right}: The variation of the anisotropy parameter, $\beta$(r) as a function of radius. They transition from more tangentially biased orbits at small radii ($<$50kpc) and become radially biased with increasing radius.}
\protect\label{fig:sig}
\end{figure*}

In Fig.\ref{fig:sig} we plot the MOND los velocity dispersion for both galaxies along with the data obtained by KP07. The fit parameters are listed in Table 1.

We found that the projected number density profile was very constraining to possible fits and if we ignored its shape at small radii as per KP07, we would require lower host galaxy masses. Remaining consistent with the projected number density, we need a mass exactly that quoted by KP07 for galaxy 1 and 40\% higher for galaxy 2, which is compatible with the uncertainties linked with the mass-to-light ratios.

Clearly, the fits require substantially radial biased orbits in the outer parts. We use the same $\beta(r)$ for both galaxies, although galaxy 2 (the more massive one) is the more constraining. Both galaxies require $\beta > 0.6$ for radii greater than 200kpc. The physics of the solutions is as follows: from the Jeans equation (Eq.1), to get a high radial dispersion, we need a low absolute value of $\gamma$ (which from Eq.1 must be a negative number in the outskirts where the variation of $\sigma_r$ is getting nearly constant, cancelling the first term in Eq.1). This can be achieved by having a low absolute value of $\alpha$, but also one tending to get steeper in the outskirts to delay the constancy of $\sigma_r$. However, the $\alpha$ needed is too low in absolute value to account for the observed slope. By taking high $\beta$, we manage to fit at the same time the slope of $\Sigma(R)$ and the los dispersion. This high radial anisotropy at large radii is exactly what is expected for self-gravitating populations from simulations of elliptical galaxy formation in MOND at radii considerably larger than the half-mass radius of the galaxy (Nipoti et al. 2007a, see their Fig.2). Although it is unclear how valid it is to expand this result to the anisotropy of the satellites test-particle population, we can a priori expect satellites to indeed have such a high radial anisotropy.

An important question is whether the distribution function (DF) corresponding to the model is positive everywhere and stable. A necessary condition is that $\gamma(r)<0$ everywhere, which is satisfied. When the spatial range of radial orbits is wide, there is a risk of radial-orbit  instability (e.g. Aguilar \& Merritt, 1990). However, this is not a concern here since the regions considered are largely outside the
self-gravitating part of the elliptical galaxy, and only satellites
taken as test particles are orbiting in this radial range. 

\section{Conclusion and Discussion}
Building upon the recent preprint of KP07, we found MOND consistently reproduces the observed declining los dispersions. By solving the Jeans equation in MOND, we have shown that even with the two very constraining datasets consisting of the projected number density profile of satellite galaxies and of their velocity dispersion profile (for a range of host galaxy masses), the data are fully consistent with MOND by simply including an increasing radial anisotropy that requires a single free parameter. These increasing anisotropies are very similar to those found by Nipoti et al. (2007a) from their simulations of dissipationless collapse in MOND. The masses of the host galaxies are reasonable and in accordance with the luminosities of the host galaxies. It is surely possible to obtain even better fits to the projected surface density and los velocity dispersions with more freedom, but the data hardly warrant it.

Finally, we comment that another preprint by Moffat \& Toth (2007, hereafter MT07) also suffers from an incomplete analysis of MOND which neglected the radial anisotropy and the correct variation of the slope of the tracer density profile, $\alpha(r)$. MT07 claim that the los velocity dispersions are in full agreement with their favoured gravity theory dubbed MOG even though they fit only the los dispersions of the more massive host satellite (galaxy 2, although in fairness this is the more difficult galaxy to use). The goodness of fit looks inferior to the MOND one presented here. Whether their theory can fit the dispersions of the lower mass host with the same set of free parameters remains to be seen. Furthermore their fit requires a host galaxy mass of $6\times10^{11}\msun$ which is more than twice our value and three times that used by KP07 and corresponds to an unlikely g-band mass-to-light ratio of 9-14.

\section*{Acknowledgements}
GWA is supported by a PPARC scholarship. BF is an FNRS research associate. GWA thanks Moti Milgrom and Bob Sanders for assisstance with the analysis. The authors wish to thank the referee, Luca Ciotti, for his valuable comments on the manuscript.
\\
The article is dedicated to the memory of Sarah Jon Main.

\label{lastpage}

\end{document}